\documentclass[5p
,number
,sort&compress
]{elsarticle}

\usepackage[%pdftex
]{graphicx}
\usepackage{dcolumn}
\usepackage{bm}
\usepackage{cases
,comment
}
\usepackage[pdftex]{hyperref}
\usepackage{color}
\definecolor{indigo(dye)}{rgb}{0.0, 0.25, 0.42}
\definecolor{phthaloblue}{rgb}{0.0, 0.06, 0.54}
\hypersetup{
    colorlinks=true,
    linkcolor=phthaloblue,
    citecolor=phthaloblue,
    filecolor=phthaloblue,
    urlcolor=phthaloblue,
    }
\usepackage[utf8]{inputenc}

\usepackage{xspace}
\usepackage{braket}

\usepackage{%amsmath,
amssymb}
\newcommand{\pdif}[2]{\frac{\partial #1}{\partial #2}}

\newcommand{\dd}{\mathrm{d}}

\newcommand{\ee}{\mathrm{e}}

\newcommand{\Mpl}{M_\mathrm{Pl}}
\newcommand{\ns}{n_{{}_\mathrm{S}}}

\newcommand{\cl}{\mathrm{cl}}
\newcommand{\sto}{\mathrm{sto}}

\newcommand{\uf}{\mathrm{f}}

\newcommand{\calL}{\mathcal{L}}

\newcommand{\calN}{\mathcal{N}}
\newcommand{\calO}{\mathcal{O}}

\newcommand{\bee}[1]{\begin{eqnarray} #1 \end{eqnarray}}

\newcommand{\bfe}[4]{
\begin{figure} 
	\centering
	\includegraphics[#1]{#2}
	\caption{#3}
	\label{#4}
\end{figure}}

\newcommand{\efolds}{$e$-folds\xspace}
\newcommand{\efolding}{$e$-folding\xspace}
\newcommand{\FP}{\mathrm{FP}}

\begin{document}
    
\title{Stochastic inflation with an extremely large number of \efolds}
\author[1,2]{Naoya Kitajima}
\ead{kitajima@tuhep.phys.tohoku.ac.jp}
\author[3]{Yuichiro Tada}
\ead{tada.yuichiro@e.mbox.nagoya-u.ac.jp}
\author[2,4]{Fuminobu Takahashi}
\ead{fumi@tohoku.ac.jp}

\address[1]{Frontier Research Institute for Interdisciplinary Sciences, Tohoku University, Sendai, 980-8578 Japan}
\address[2]{Department of Physics, Tohoku University, Sendai, 980-8578 Japan}
\address[3]{Department of Physics, Nagoya University, Nagoya 464-8602, Japan}
\address[4]{Kavli IPMU (WPI), UTIAS, The University of Tokyo, Kashiwa, Chiba 277-8583, Japan}

\begin{abstract}
We propose a class of single-field, slow-roll inflation models in which a typical number of \efolds can be extremely large. The key point is to introduce a very shallow local minimum near the top of the potential in a hilltop inflation model. In particular, a typical number of \efolds is 
enhanced if classical behavior
dominates around the local minimum such that the inflaton probability distribution is drifted to the local minimum as a whole. After the inflaton escapes from the local minimum due to the stochastic dynamics, the ordinary slow-roll inflation follows and it can generate the primordial density perturbation consistent with observation. 
Interestingly, our scenario inherits
the advantages of the old and new inflation: the typical \efolds can be extremely large as in the old inflation, and slow-roll inflation
naturally follows after the stochastic regime as in the new inflation.
In our numerical example, the typical number of \efolds can be as large as $10^{10^{10}}$, 
which is large enough for various light scalars such the QCD axion to reach the Bunch-Davies distribution. 
\end{abstract}


\date{\today}
\maketitle

%%%%%%%%%%%%%%%%%%%%%%%%%%%%%%%%%%%%%%%%
\section{Introduction}
\label{sec:introduction}
%%%%%%%%%%%%%%%%%%%%%%%%%%%%%%%%%%%%%%%%

How long can inflation last? From the observational point of view, the total number of \efolds, $N$, must be larger than $\sim 50$--$60$, which 
weakly depends on the inflation scale and thermal history after inflation. On the other hand, an extremely long duration of inflation is often required in certain scenarios. For instance, the relaxion model~\cite{Graham:2015cka}, the stochastic axion 
scenario~\cite{Graham:2018jyp,Guth:2018hsa,Ho:2019ayl,Takahashi:2019pqf}, a quintessence model~\cite{Ringeval:2010hf,Ringeval:2019bob} demand the \efolding number of $\log_{10}N\sim\calO(10)$, depending on model parameters.\footnote{
If the \efolding number is extremely large, those scenarios that involve scalars with fine-tuned initial conditions
might be significantly modified. For example, such a problem in the curvaton paradigm is reported in Ref.~\cite{Torrado:2017qtr}.
Also, rare events such as the Higgs tunneling to large-field values could take place somewhere in the entire universe
(see e.g. Ref.~\cite{Enqvist:2014bua}).}
The purpose of this Letter is to provide a simple single-field, slow-roll inflation model whose {\it typical} number of \efolds is extremely large.

There is a variety of inflation models which last very long. In the string/axion 
landscape~\cite{Bousso:2000xa,Susskind:2003kw,Freivogel:2005vv,Higaki:2014pja,Higaki:2014mwa,Wang:2015rel,Masoumi:2016eqo,Bachlechner:2019vcb}, there are many local minima where the old inflation takes place and continues until the inflaton tunnels toward one of the adjacent local minima with a lower energy
through bubble formation. 
In this case, the typical \efolding number can be exponentially large. However, one needs slow-roll inflation after the bubble formation to explain the observed cosmic microwave background (CMB) temperature/polarization anisotropies. For this, one may need  another light scalar, in which case 
the two inflation scales are not related to each other, in general.

It is also well known that a large class of inflation models can be eternal~\cite{Linde:1982ur,Steinhardt:1982kg,Vilenkin:1983xq,Linde:1986fc,Linde:1986fd,Goncharov:1987ir} (see also \cite{Guth:2000ka,Guth:2007ng,Linde:2015edk}).
The stochastic dynamics of the inflaton plays a crucial role in eternal inflation: quantum fluctuations of the inflaton drive it upward or downward
compared to the classical motion. In particular, upward quantum fluctuations keep the inflaton from rolling
down the potential, leading to eternal inflation.\footnote{
We note that the downward fluctuations help the inflaton to escape from the eternal inflation regime.}
In eternal chaotic inflation~\cite{Linde:1986fc,Linde:1986fd}, quantum fluctuations overcome the classical motion at 
sufficiently large fields values, typically much larger than the Planck scale.
Eternal inflation also occurs  in the new or hilltop inflation~\cite{Linde:1982ur,Steinhardt:1982kg,Vilenkin:1983xq}. This is because the inflaton dynamics becomes stochastic in the vicinity of the potential maximum where the classical motion is suppressed. 
In both cases, once the universe enters the stochastic regime,  
the inflation continues and never ends in some regions.
The volume of the universe is dominated by such regions where the inflation continues. In this sense the inflation is eternal.

It is worth noting that the eternal inflation is based on the idea that the inflating region expands and dominates the volume at later times.
In fact, if one randomly picks up a point in space where the inflaton dynamics is in the stochastic regime, 
the expected \efolds there is not infinite but finite~\cite{Barenboim:2016mmw} (see also Ref.~\cite{Assadullahi:2016gkk}).
In other words, one needs to 
choose a specific point in space to have a sufficiently long inflation. Such a special 
choice or fine-tuning can be compensated by large physical volume due to the prolonged inflation. Thus, the eternity of the eternal inflation relies
on the volume measure. 

The purpose of this Letter is to show that one can make the {\it typical} number of \efolds extremely large
in a context of a simple single-field, slow-roll inflation. In particular, it does not rely on the volume measure. 
In Sec.~\ref{sec:stochastic} we address our scenario after a brief review on the stochastic formalism.
In Sec.~\ref{sec:model} we provide an example model which exhibits an extremely long inflation. 
The last section is devoted for discussion and conclusions.

%%%%%%%%%%%%%%%%%%%%%%%%%%%%%%%%%%%%%%%%
\section{Stochastic inflation with large $N$}
\label{sec:stochastic}

In the simplest setup, the inflationary phase in the early universe can be realized by the potential energy of some scalar field inflaton, $\phi$, homogeneously filling the universe. For a sufficiently long inflationary phase, however,
the inflaton field must be extremely light, in which case its homogeneity beyond the horizon scale is no longer guaranteed.
Specifically, superhorizon inhomogeneities are sourced by subhorizon fluctuations which continuously exit the horizon during inflation. 
Even though the subhorizon fluctuations originate from the quantum zero-point oscillation, they are well approximated
to be classical ones after the horizon exit due to the gravitational decoherence.
Accordingly the evolution of the local inflaton field value can be described by a diffusion process. This is known as the stochastic approach~\cite{Starobinsky:1986fx,Starobinsky:1994bd}, in which the probability distribution function $P(\phi)$ follows the Fokker-Planck (FP) equation,
\bee{
    \pdif{P}{N}=\calL_\FP(\phi)\cdot P, \quad \calL_\FP=-\pdif{}{\phi}h(\phi)+\frac{1}{2}\pdif{^2}{\phi^2}D(\phi).
}
Here $N$ is the \efolding number used as a time variable, and it is related to the  Hubble parameter $H$ and the cosmic time $t$
by $\dd N=H\dd t$.
For single-field and slow-roll inflation, the drift and diffusion coefficients are given by $h=-\Mpl^2V^\prime/V$ and $D=V/12\pi^2\Mpl^2$ 
in terms of the inflaton potential $V(\phi)$, respectively.
Due to such a random walk behavior, the \efolds $\calN(\phi)$ elapsed for the first passage from an initial field value $\phi$ 
to $\phi_\uf$ at the end of inflation also becomes a random parameter. 
Its generating function $\chi_\calN(\phi;J)=\braket{\ee^{i J\calN(\phi)}}$ is known to follow 
the adjoint FP equation~\cite{Vennin:2015hra,Pattison:2017mbe},
\bee{
    \calL_\FP^\dagger\cdot\chi_\calN=-iJ\chi_\calN, \quad \calL_\FP^\dagger=h(\phi)\pdif{}{\phi}+\frac{1}{2}D(\phi)\pdif{^2}{\phi^2},
}
with the boundary condition $\chi_\calN=1$ at the end of inflation.
The bracket represents an average over realizations.
An arbitary moment of $\calN$ is obtained from this generating function by $\braket{\calN^n}=\left.\left(\frac{1}{i}\pdif{}{J}\right)^n\chi_\calN\right|_{J=0}$.
Thus, one can estimate the \emph{typical} \efolding number of inflation by calculating
the average of the elapsed \efolds, $\braket{\calN}$.

For the single-field slow-roll inflation, these FP equations are reduced to the ordinary differential equations, for which
the moments of $\calN$ have recursive analytic solutions. For example, the expectation value of $\calN$ is given by
\bee{\label{eq: calN}
    \braket{\calN}(\phi)=\int^\phi_{\phi_\uf}\frac{\dd x}{\Mpl}\int^{\bar{\phi}}_x\frac{\dd y}{\Mpl}\frac{1}{v(y)}\exp\left[\frac{1}{v(y)}-\frac{1}{v(x)}\right], 
}
where $v=\frac{V}{24\pi^2\Mpl^4}$ is the normalized potential in the Planck unit, and $\bar{\phi}$ represents a constant of integration 
at which the first derivative vanishes $\partial_\phi\braket{\calN}|_{\bar{\phi}}=0$. For a $Z_2$-symmetric hilltop model we consider later, 
$\bar{\phi}=0$ is obviously suitable for this boundary condition.
As shown in Ref.~\cite{Vennin:2015hra},
when the following conditions are satisfied,
\bee{\label{eq: classicality condition}
    v\ll1 \quad \mathrm{and} \quad \eta_\cl=\left|\frac{v^{\prime\prime}v^2}{{v^\prime}^2}\right|\ll1,
}
one can use the saddle-point approximation to simplify 
this analytic solution (\ref{eq: calN}) 
 to the well-known formula for the classical drift-dominated case,
\bee{\label{eq: Ncl}
    \braket{\calN}(\phi)\simeq N_\cl(\phi)=\int^\phi_{\phi_\uf}\frac{\dd x}{\Mpl^2}\frac{v(x)}{v^\prime(x)}.
}
Note that this formula is valid as long as the conditions (\ref{eq: classicality condition}) 
are satisfied everywhere between $\phi$ and $\phi_\uf$. 
Since the first condition in Eq.~(\ref{eq: classicality condition}) is trivially satisfied in our context, 
it is the classicality parameter $\eta_\cl$ that controls the stochastic effect on the averaged \efolding number.

For a better understanding of the second condition in Eq.~(\ref{eq: classicality condition}), let us consider the change of the inflaton potential value due to the inflaton dynamics in one Hubble time: $\delta v \simeq v' \delta\phi + \frac{1}{2}v'' \delta\phi^2$, where $\delta\phi$ is the sum of the classical motion $\delta\phi_\cl \sim h = -\Mpl^2 v'/v$ and the quantum jump $\delta\phi_\sto \sim \sqrt{D} = \sqrt{2v}\Mpl$. Note that the ensemble average and the variance of $\delta\phi$ are respectively given by $\langle \delta \phi\rangle = \delta\phi_\cl$ and $\langle \delta\phi^2 \rangle = \delta\phi_\cl^2 + \delta\phi_\sto^2$, and hence one obtains $\langle \delta v \rangle \approx v' \delta\phi_\cl + \frac{1}{2}v''(\delta\phi_\cl^2+\delta\phi_\sto^2)$. Thus, the second condition in Eq.~(\ref{eq: classicality condition}) requires that the averaged-change of the inflaton potential value is dominated by the linear term of the classical motion, i.e. $|v' \delta\phi_{\cl}| \gg |v''\delta\phi_\sto^2|$. 
Note that $v'\delta\phi_\cl \gg v''\delta\phi_\cl^2$ is satisfied under the second slow-roll condition: $\eta_V \ll 1$ with the second slow-roll parameter $\eta_V=\Mpl^2v^{\prime\prime}/v$.

Let us  emphasize here that the above classical formula~(\ref{eq: Ncl}) can be valid 
even if there is a highly stochastic region between $\phi$ and $\phi_\uf$, 
as long as the curvature of the potential is sufficiently small. 
To see this, let us first define the stochasticity of the inflaton dynamics by $\xi_\sto\equiv\delta\phi_\sto^2/\delta\phi_\cl^2\,(\equiv h^2/D) = 2v^3/{v^\prime}^2\Mpl^2$, which 
is nothing but the amplitude of the curvature perturbations in the classical limit. For $\xi_\sto \gtrsim 1$, 
the size of the quantum jump is greater than the classical field excursion over one Hubble time, i.e. the inflaton
dynamics is in the stochastic regime (and this is nothing but the necessary condition for eternal inflation). 
The stochasticity parameter is related to $\eta_\cl$ by $\eta_\cl=\frac{1}{2}|\eta_V|\xi_\sto$.
Therefore, the conditions (\ref{eq: classicality condition}) can be satisfied in a highly stochastic region $\xi_\sto \gtrsim 1$
if $\eta_V\ll1$, in which case the expected \efolds can be well approximated by the classical formula~(\ref{eq: Ncl}).

As an extreme example,
let us consider a slightly tilted linear potential for which $\xi_\sto \gtrsim 1$ and $\eta_\cl$ vanishes
by definition. Suppose that the inflaton is initially located at some point.
Then, the inflaton distribution spreads out due to the quantum diffusion
because of $\xi_\sto \gtrsim 1$. 
However, the quantum diffusion does not change the inflaton potential value on average
because up and downhills occur with the same probability. It is the classical motion that
changes the potential value on average, as it uniformly drifts the whole inflaton distribution.
 
If $\eta_\cl \gtrsim 1$, 
the diffusion dynamics overcomes the uniform drift due to the classical motion. In other words, the evolution of the universe
relies upon the stochastic process, and the classical dynamics is irrelevant. In the usual hilltop model (without a local minimum), 
$\eta_\cl$ is large around the potential maximum, and the stochastic diffusion sweeps out the inflaton from the hilltop. This is
the reason why the averaged \efolds are saturated and does not grow as the initial value $\phi$ approaches the hilltop (see Fig.~\ref{fig: Nvsphi}).

In fact, one can slightly modify the hilltop potential to realize an extremely large \efolds. 
To this end, we introduce a shallow local minimum surrounded by a region with $\eta_\cl \ll 1$ around the hilltop. 
There, the inflaton distribution tends to be pushed back to the local minimum as a whole even in the stochastic regime,
and then the expected \efolds can be significantly enhanced.  
In such a case, the escape rate can be analytically estimated by the expansion around the local maximum and minimum for $x$ and $y$ integrals respectively in Eq.~(\ref{eq: calN}), as given by~\cite{Noorbala:2018zlv}
\bee{\label{eq: Nesc}
    \braket{\calN}_\mathrm{esc}\simeq\frac{\pi v(\phi_+)}{2\Mpl^2\sqrt{v^{\prime\prime}(0)|v^{\prime\prime}(\phi_+)|}}\ee^{\frac{1}{v(0)}-\frac{1}{v(\phi_+)}},
}
for a symmetric potential. Here $\phi_+$ denotes the local maximum and we take the local minimum at the origin $\phi=0$. As suggested by the Hawking-Moss factor $\ee^{1/v(0)-1/v(\phi_+)}$~\cite{Hawking:1981fz}, this escape time can be extremely large for a sufficient low scale potential $v\ll1$. In the next section, we concretely provide such a model which generates the primordial density perturbations 
consistent with the CMB observations after exiting the local minimum.

%%%%%%%%%%%%%%%%%%%%%%%%%%%%%%%%%%%%%%%%

%%%%%%%%%%%%%%%%%%%%%%%%%%%%%%%%%%%%%%%%
\section{Example model}
\label{sec:model}

Let us explicitly show a concrete model of the extreme long inflation. We consider a
 $Z_2$-symmetric hilltop model given by
\bee{
\label{vphi}
    V(\phi)=\frac{1}{2}m^2\phi^2-\frac{1}{4}\lambda\phi^4+\left(\Lambda^2-g\frac{\phi^6}{\Mpl^4}\right)^2,
}
where $\Lambda$ determines the inflation scale, $\lambda$ and $g$ are positive coupling constants, and
$m$ is the mass at the origin. 
For a positive mass-squared in the range of $0<m^2 < H^2$, this potential has a shallow local minimum around the origin $\phi=0$, 
where the inflaton spends a large number of \efolds. After diffusing out from the shallow local minimum, 
the inflaton dynamics is driven by the tilt of the quartic and hexic terms as an ordinary hilltop inflation,
toward the true minimum at $\phi_\mathrm{min}\simeq(\Lambda^2\Mpl^4/g)^{1/6}$.
Some of the parameters are fixed by the CMB data.
For example, the following parameters
\bee{\label{eq: params}
	&&\left(\frac{\Lambda}{\Mpl},\,\lambda,\,\left(\frac{\Lambda^2\Mpl^4}{g}\right)^{1/6}\right) \nonumber \\
	&&\qquad=\left(0.998\times10^{-4},\,3.39\times10^{-14},\,0.1\Mpl\right),
}
result in the curvature perturbations with the amplitude $A_s=2.1\times10^{-9}$ and spectral index $\ns=0.958$ 
consistent with the recent Planck observation~\cite{Aghanim:2018eyx}.\footnote{
By adding a linear term~\cite{Takahashi:2013cxa} or the Coleman-Weinberg correction~\cite{Nakayama:2012dw},
one can increase $n_s$ to give a better fit to observation. 
}
We illustrate the potential form in Fig.~\ref{fig: potential} with these parameters and the typical mass value $m^2/H^2=10^{-7}$.
Here $H=\Lambda^2/\sqrt{3}\Mpl$ is the Hubble parameter around the origin.
For this or larger mass, the local minimum becomes deep enough so that the \emph{classical} region with $\eta_\cl < 1$ 
appears between the local minimum and maximum. The classical region efficiently pushes the inflaton back to the local minimum 
and the expected \efolds will be enhanced significantly as shown below.
The stochasticity itself is high enough as $\xi_\sto>1$ for $\phi\lesssim10^{-4}\Mpl$ enclosing the whole local minimum and maximum.

\bfe{width=\hsize}{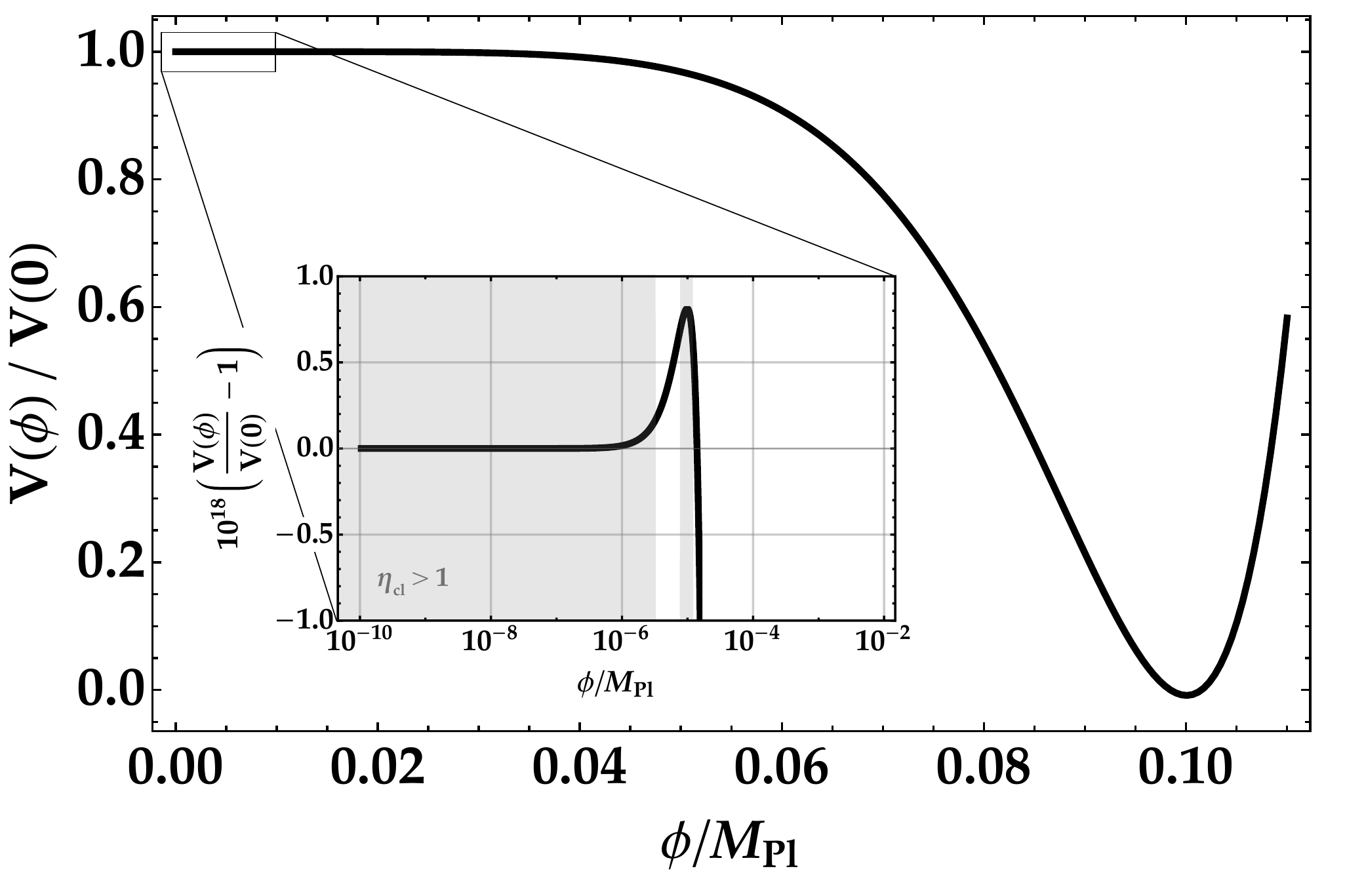}{The potential~(\ref{vphi}) with the parameter values~(\ref{eq: params}) and the typical mass $m^2/H^2=10^{-7}$.
The potential has a shallow local minimum surrounded by the local maximum as
shown in the magnification. The shaded region corresponds with $\eta_\cl>1$. The \emph{classical} region with $\eta_\cl<1$ appears 
between the local minimum and maximum that will increase the expected \efolds for $m^2/H^2 \gtrsim 10^{-7}$, because the inflaton is pushed
back to the local minimum by the classical dynamics.}{fig: potential}

Fig.~\ref{fig: Nvsphi} shows the results of the numerical integration of $\braket{\calN}(\phi)$ in Eq.~(\ref{eq: calN}) 
for different masses. One can see that, at sufficiently small $\phi$,
the expected \efolds is saturated due to the stochastic effects, which sweep out the inflaton from the hilltop, $\phi\lesssim10^{-5}\Mpl$.
For $\phi\gtrsim10^{-5}\Mpl$, the inflaton already passes to the right of the local maximum (in the case of $m^2 > 0$),
and  $\braket{\calN}$ is reduced to the classical formula $N_\cl$ in Eq.~(\ref{eq: Ncl}) even though 
the stochasticity $\xi_\sto$ itself is large until around $\phi\lesssim10^{-4}\Mpl$. 
The usual standard slow-roll inflation will take place afterwards, generating the primordial density perturbations consistent with the CMB observation.
There is a slight wiggling around $\phi\sim10^{-2}\Mpl$, which, however, is caused by the numerical error, and not due to any physical effect. 
The expected \efolds become enhanced for $m^2/H^2\gtrsim10^{-7}$ as mentioned.

\bfe{width=\hsize}{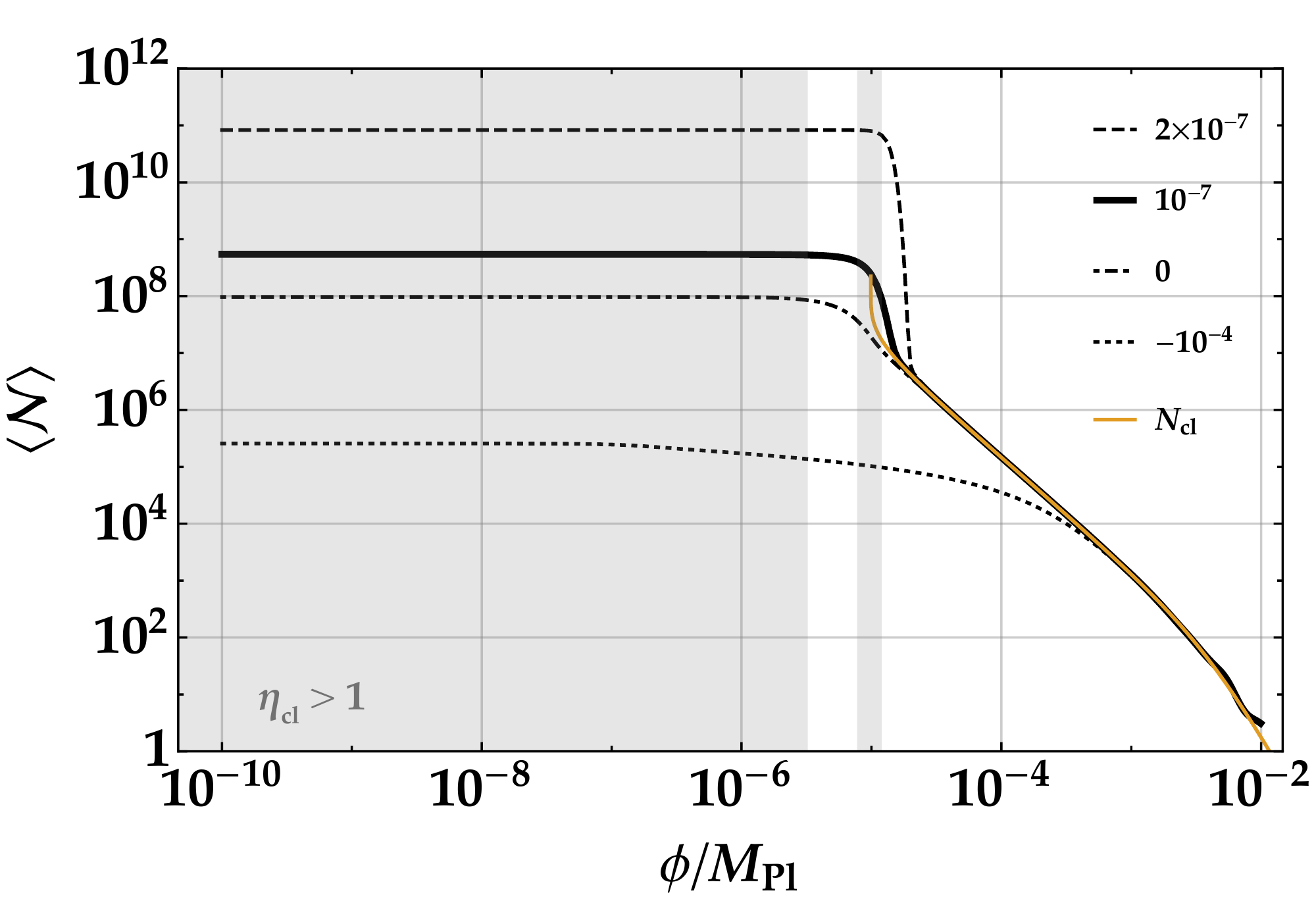}{The averaged \efolds $\braket{\calN}(\phi)$ given by Eq.~(\ref{eq: calN}) 
for different masses, $m^2/H^2=(2\times10^{-7},\,10^{-7},\,0,\,-10^{-4})$, where $H=\Lambda^2/\sqrt{3}\Mpl$ is the Hubble parameter around the origin. The other parameters are fixed by the CMB data as Eq.~(\ref{eq: params}). The shaded region and the orange line represent $\eta_\cl>1$ and $N_\cl(\phi)$ given by Eq.~(\ref{eq: Ncl}), respectively,
for $m^2/H^2=10^{-7}$. 
For $\phi\gtrsim10^{-5}\Mpl$, the inflaton is to the right of the local maximum, and $\braket{\calN}$ is reduced to the classical formula $N_\cl$ without the diffusion.  Then, a sufficiently long standard slow-roll inflation follows. 
For $m^2/H^2\gtrsim10^{-7}$ the expected \efolds are enhanced due to the \emph{classical} region with $\eta_\cl<1$ between the local minimum and maximum.}{fig: Nvsphi}

We show in Fig.~\ref{fig: meanN}  the $m^2$-dependence of the \efolds from the origin $\phi=0$. We have confirmed that the analytic estimate of the escape time~(\ref{eq: Nesc}) agrees well with our numerical result.
One can see that an extremely large number of \efolds as $\braket{\calN}\sim10^{10^{10}}$ can be realized with a deep enough local minimum. 
Even in such a case, the inflaton mass-squared itself is still smaller than the Hubble scale, so that the slow-roll stochastic approach remains valid. 
The typical \efolds in this model is large enough for various light scalars such the QCD axion to reach the Bunch-Davies distribution~\cite{Bunch:1978yq}.
For instance, the required \efolds 
for the axion to reach the so-called Bunch-Davies distribution
is $N \gtrsim 10^{26} (H_{\rm inf}/100 {\rm MeV})^2/(m_a/10^{-5}{\rm eV})^2$~\cite{Graham:2018jyp,Guth:2018hsa}.\footnote{
The typical \efolds can be even larger if one considers a lower inflation scale. 
}
For comparison, 
we also show the result for the usual case with the negative mass-squared, where
the averaged \efolding number decreases as expected.

\bfe{width=\hsize}{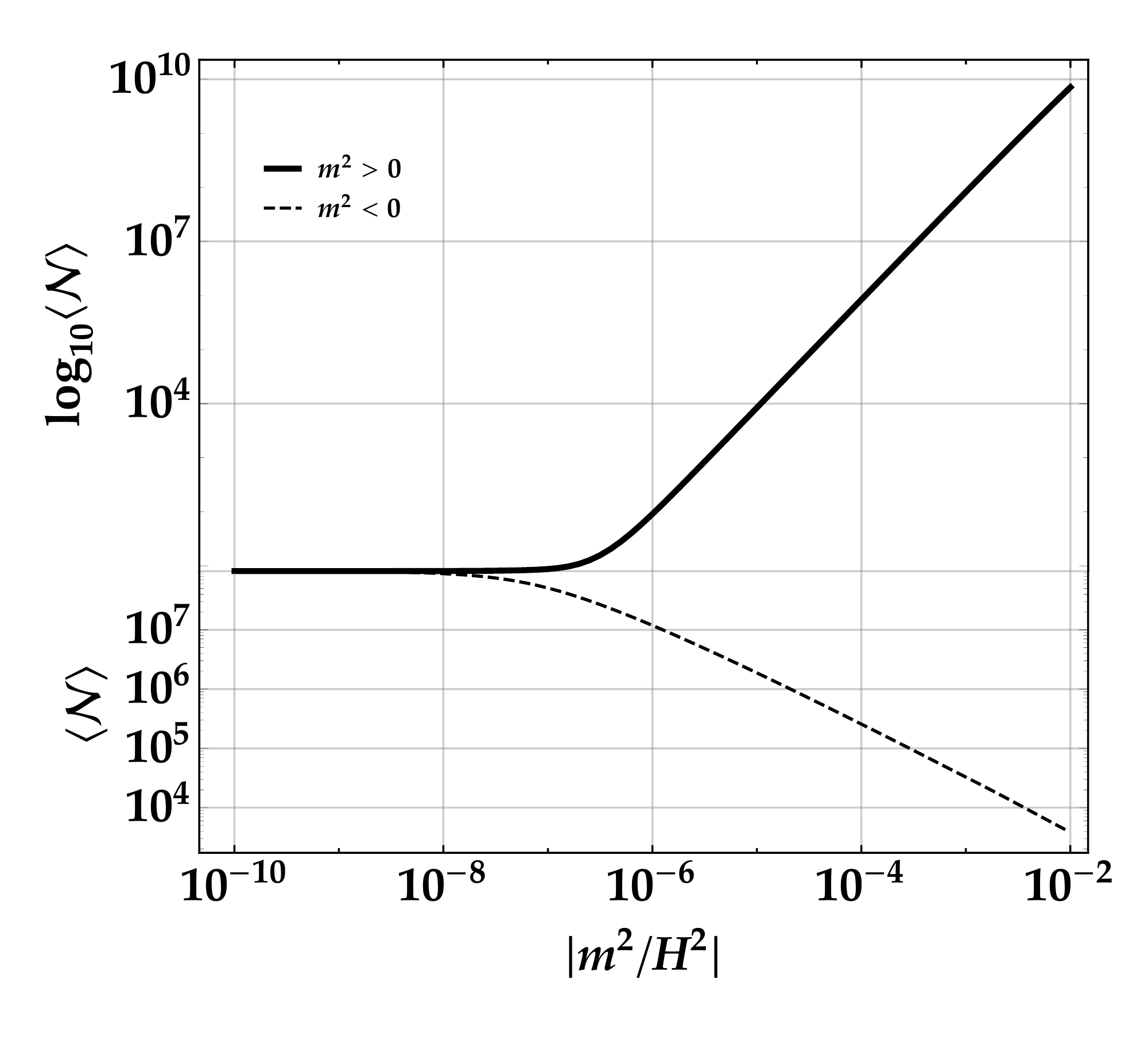}{The dependence of $\braket{\calN}(\phi=0)$ on $m^2$. Again the other parameters are given by Eq.~(\ref{eq: params}).
For sufficiently large and positive mass squared (solid line)
(but still smaller than $H^2$), extremely long inflation is realized. Typical \efolds 
can be as large as $\braket{\calN}\sim10^{10^{10}}$.
The numerical result is consistent with the analytic estimate~(\ref{eq: Nesc}).
The case of the negative mass squared (dashed line) is also shown for comparison.
}{fig: meanN}

%%%%%%%%%%%%%%%%%%%%%%%%%%%%%%%%%%%%%%%%

%%%%%%%%%%%%%%%%%%%%%%%%%%%%%%%%%%%%%%%%
\section{Discussion and Conclusions}
\label{sec:conculsion}

So far we have studied a simple hilltop inflation with a shallow local minimum at the origin. One can realize a hilltop inflation
in terms of an axion field which enjoys discrete symmetry~\cite{Czerny:2014wza,Czerny:2014xja}. In this case
a small modulation(s) can be introduced, which may induce multiple local minima around the hilltop. See Refs.~\cite{Flauger:2009ab,Kobayashi:2010pz,Takahashi:2013tj} for cosmological implications of such small modulations on the inflaton potential.
More comprehensive study for wider classes of the inflaton potential is left for future work.

If $m^2$ is taken to be larger than $H^2$ in the inflation model (\ref{vphi}), the local minimum becomes deeper, and
the stochastic formalism breaks down.
The inflaton trapped in the local minimum may tunnel to the right of the potential barrier by the bubble formation, followed by
slow-roll inflation as in the case of the original new inflation~\cite{Linde:1981mu,Albrecht:1982wi}. For even larger $m^2$, the situation
will be similar to the old inflation~\cite{Guth:1980zm,Starobinsky:1980te,Sato:1980yn}. In this sense, our scenario corresponds to
a limit of the shallow local minimum for which the stochastic formalism can be applied. Interestingly, our scenario inherits
the advantages of the old and new inflation: the typical \efolds can be extremely large as in the old inflation, and slow-roll inflation
naturally follows afterward as in the new inflation.

In this Letter, we have studied a possibility to realize extremely long inflation in the simple single-field slow-roll framework. 
In particular we work on a small-field inflation, which can be well described by an effective field theory.
In the usual eternal inflation picture,  the quantum diffusion dominates over the classical dynamics, 
and it keeps the inflaton from rolling down the potential in some spatial regions. The inflation continues in such regions, 
which subsequently dominate the volume of the universe even if the probability to continue inflation is small. 
On the other hand, we have explored a possibility to realize  an extremely large number of \emph{typical} \efolds 
$\braket{\calN}$, i.e. the expected \efolding number without resort to the volume measure. We have shown that this is
indeed possible by introducing a shallow local minimum around the hilltop, which was numerically confirmed using a 
simple toy model: see Fig.~\ref{fig: meanN} showing the dependence of $\braket{\calN}$ on the curvature of the local minimum, which 
can be understood as the Hawking-Moss decay rate~(\ref{eq: Nesc})~\cite{Hawking:1981fz,Noorbala:2018zlv}.
It is the classical region with $\eta_\cl<1$ to the left of the local maximum that pushes back the inflaton to the local minimum,
and forces the inflaton to stay there for an extremely long time. Although extremely unlikely, it is possible for the inflaton to
move to the right of the local maximum due to the accumulated quantum diffusion. Then, 
after moving to the right of the local maximum, standard slow-roll inflation follows as shown in Fig.~\ref{fig: Nvsphi} and 
generates the primordial density perturbation consistent with observation.

We emphasize that our scenario does not rely on the volume measure and thus does not need any fine-tuning of the spatial position to
realize extremely long inflation.
Moreover, the eternal inflation and the slow-roll inflation phases are smoothly connected in a single-field regime, without any processes of 
tunneling and bubble nucleation.
Therefore, it provides an alternative possibility to naturally realize extremely large number of \efolds, which is often necessary in
some cosmological scenarios.

%%%%%%%%%%%%%%%%%%%%%%%%%%%%%%%%%%%%%%%%

%%%%%%%%%%%%%%%%%%%%%%%%%%%%%
\section*{Acknowledgments}
F.T. thanks Wen Yin for discussion on the eternal inflation.
This work is supported by JSPS KAKENHI Grant Numbers
JP15H05889 (F.T.), JP15K21733 (F.T.),  JP17H02875 (F.T.),
JP17H02878 (F.T.), JP18J01992 (Y.T.), JP19K14707 (Y.T.), JP18H01243 (N.K.), JP19K14708 (N.K.), JP19H01894 (N.K.), and by World Premier International Research Center Initiative (WPI Initiative), MEXT, Japan.
%%%%%%%%%%%%%%%%%%%%%%%%%%%%%

\nocite{}
\bibliographystyle{elsarticle-num}
\bibliography{main}

\begin{thebibliography}{10}
\expandafter\ifx\csname url\endcsname\relax
  \def\url#1{\texttt{#1}}\fi
\expandafter\ifx\csname urlprefix\endcsname\relax\def\urlprefix{URL }\fi
\expandafter\ifx\csname href\endcsname\relax
  \def\href#1#2{#2} \def\path#1{#1}\fi

\bibitem{Graham:2015cka}
P.~W. Graham, D.~E. Kaplan, S.~Rajendran, {Cosmological Relaxation of the
  Electroweak Scale}, Phys. Rev. Lett. 115~(22) (2015) 221801.
\newblock \href {http://arxiv.org/abs/1504.07551} {\path{arXiv:1504.07551}},
  \href {https://doi.org/10.1103/PhysRevLett.115.221801}
  {\path{doi:10.1103/PhysRevLett.115.221801}}.

\bibitem{Graham:2018jyp}
P.~W. Graham, A.~Scherlis, {Stochastic axion scenario}, Phys. Rev. D98~(3)
  (2018) 035017.
\newblock \href {http://arxiv.org/abs/1805.07362} {\path{arXiv:1805.07362}},
  \href {https://doi.org/10.1103/PhysRevD.98.035017}
  {\path{doi:10.1103/PhysRevD.98.035017}}.

\bibitem{Guth:2018hsa}
F.~Takahashi, W.~Yin, A.~H. Guth, {QCD axion window and low-scale inflation},
  Phys. Rev. D98~(1) (2018) 015042.
\newblock \href {http://arxiv.org/abs/1805.08763} {\path{arXiv:1805.08763}},
  \href {https://doi.org/10.1103/PhysRevD.98.015042}
  {\path{doi:10.1103/PhysRevD.98.015042}}.

\bibitem{Ho:2019ayl}
S.-Y. Ho, F.~Takahashi, W.~Yin, {Relaxing the Cosmological Moduli Problem by
  Low-scale Inflation}, JHEP 04 (2019) 149.
\newblock \href {http://arxiv.org/abs/1901.01240} {\path{arXiv:1901.01240}},
  \href {https://doi.org/10.1007/JHEP04(2019)149}
  {\path{doi:10.1007/JHEP04(2019)149}}.

\bibitem{Takahashi:2019pqf}
F.~Takahashi, W.~Yin, {QCD Axion on Hilltop by Phase Shift of $\pi$} (2019).
\newblock \href {http://arxiv.org/abs/1908.06071} {\path{arXiv:1908.06071}}.

\bibitem{Ringeval:2010hf}
C.~Ringeval, T.~Suyama, T.~Takahashi, M.~Yamaguchi, S.~Yokoyama, {Dark energy
  from primordial inflationary quantum fluctuations}, Phys. Rev. Lett. 105
  (2010) 121301.
\newblock \href {http://arxiv.org/abs/1006.0368} {\path{arXiv:1006.0368}},
  \href {https://doi.org/10.1103/PhysRevLett.105.121301}
  {\path{doi:10.1103/PhysRevLett.105.121301}}.

\bibitem{Ringeval:2019bob}
C.~Ringeval, T.~Suyama, M.~Yamaguchi, {Large mass hierarchy from a small
  nonminimal coupling}, Phys. Rev. D99~(12) (2019) 123524.
\newblock \href {http://arxiv.org/abs/1903.03544} {\path{arXiv:1903.03544}},
  \href {https://doi.org/10.1103/PhysRevD.99.123524}
  {\path{doi:10.1103/PhysRevD.99.123524}}.

\bibitem{Torrado:2017qtr}
J.~Torrado, C.~T. Byrnes, R.~J. Hardwick, V.~Vennin, D.~Wands, {Measuring the
  duration of inflation with the curvaton}, Phys. Rev. D98~(6) (2018) 063525.
\newblock \href {http://arxiv.org/abs/1712.05364} {\path{arXiv:1712.05364}},
  \href {https://doi.org/10.1103/PhysRevD.98.063525}
  {\path{doi:10.1103/PhysRevD.98.063525}}.

\bibitem{Enqvist:2014bua}
K.~Enqvist, T.~Meriniemi, S.~Nurmi, {Higgs Dynamics during Inflation}, JCAP
  1407 (2014) 025.
\newblock \href {http://arxiv.org/abs/1404.3699} {\path{arXiv:1404.3699}},
  \href {https://doi.org/10.1088/1475-7516/2014/07/025}
  {\path{doi:10.1088/1475-7516/2014/07/025}}.

\bibitem{Bousso:2000xa}
R.~Bousso, J.~Polchinski, {Quantization of four form fluxes and dynamical
  neutralization of the cosmological constant}, JHEP 06 (2000) 006.
\newblock \href {http://arxiv.org/abs/hep-th/0004134}
  {\path{arXiv:hep-th/0004134}}, \href
  {https://doi.org/10.1088/1126-6708/2000/06/006}
  {\path{doi:10.1088/1126-6708/2000/06/006}}.

\bibitem{Susskind:2003kw}
L.~Susskind, {The Anthropic landscape of string theory}, in: {Carr, Bernard
  (ed.): Universe or multiverse?}, 2003, pp. 247--266.
\newblock \href {http://arxiv.org/abs/hep-th/0302219}
  {\path{arXiv:hep-th/0302219}}.

\bibitem{Freivogel:2005vv}
B.~Freivogel, M.~Kleban, M.~Rodriguez~Martinez, L.~Susskind, {Observational
  consequences of a landscape}, JHEP 03 (2006) 039.
\newblock \href {http://arxiv.org/abs/hep-th/0505232}
  {\path{arXiv:hep-th/0505232}}, \href
  {https://doi.org/10.1088/1126-6708/2006/03/039}
  {\path{doi:10.1088/1126-6708/2006/03/039}}.

\bibitem{Higaki:2014pja}
T.~Higaki, F.~Takahashi, {Natural and Multi-Natural Inflation in Axion
  Landscape}, JHEP 07 (2014) 074.
\newblock \href {http://arxiv.org/abs/1404.6923} {\path{arXiv:1404.6923}},
  \href {https://doi.org/10.1007/JHEP07(2014)074}
  {\path{doi:10.1007/JHEP07(2014)074}}.

\bibitem{Higaki:2014mwa}
T.~Higaki, F.~Takahashi, {Axion Landscape and Natural Inflation}, Phys. Lett.
  B744 (2015) 153--159.
\newblock \href {http://arxiv.org/abs/1409.8409} {\path{arXiv:1409.8409}},
  \href {https://doi.org/10.1016/j.physletb.2015.03.052}
  {\path{doi:10.1016/j.physletb.2015.03.052}}.

\bibitem{Wang:2015rel}
G.~Wang, T.~Battefeld, {Vacuum Selection on Axionic Landscapes}, JCAP 1604~(04)
  (2016) 025.
\newblock \href {http://arxiv.org/abs/1512.04224} {\path{arXiv:1512.04224}},
  \href {https://doi.org/10.1088/1475-7516/2016/04/025}
  {\path{doi:10.1088/1475-7516/2016/04/025}}.

\bibitem{Masoumi:2016eqo}
A.~Masoumi, A.~Vilenkin, {Vacuum statistics and stability in axionic
  landscapes}, JCAP 1603~(03) (2016) 054.
\newblock \href {http://arxiv.org/abs/1601.01662} {\path{arXiv:1601.01662}},
  \href {https://doi.org/10.1088/1475-7516/2016/03/054}
  {\path{doi:10.1088/1475-7516/2016/03/054}}.

\bibitem{Bachlechner:2019vcb}
T.~C. Bachlechner, K.~Eckerle, O.~Janssen, M.~Kleban, {The Axidental Universe}
  (2019).
\newblock \href {http://arxiv.org/abs/1902.05952} {\path{arXiv:1902.05952}}.

\bibitem{Linde:1982ur}
A.~D. Linde, {NONSINGULAR REGENERATING INFLATIONARY UNIVERSE} (1982).

\bibitem{Steinhardt:1982kg}
P.~J. Steinhardt, {NATURAL INFLATION}, in: {Nuffield Workshop on the Very Early
  Universe Cambridge, England, June 21-July 9, 1982}, 1982, pp. 251--266.

\bibitem{Vilenkin:1983xq}
A.~Vilenkin, {The Birth of Inflationary Universes}, Phys. Rev. D27 (1983) 2848.
\newblock \href {https://doi.org/10.1103/PhysRevD.27.2848}
  {\path{doi:10.1103/PhysRevD.27.2848}}.

\bibitem{Linde:1986fc}
A.~D. Linde, {ETERNAL CHAOTIC INFLATION}, Mod. Phys. Lett. A1 (1986) 81.
\newblock \href {https://doi.org/10.1142/S0217732386000129}
  {\path{doi:10.1142/S0217732386000129}}.

\bibitem{Linde:1986fd}
A.~D. Linde, {Eternally Existing Selfreproducing Chaotic Inflationary
  Universe}, Phys. Lett. B175 (1986) 395--400.
\newblock \href {https://doi.org/10.1016/0370-2693(86)90611-8}
  {\path{doi:10.1016/0370-2693(86)90611-8}}.

\bibitem{Goncharov:1987ir}
A.~S. Goncharov, A.~D. Linde, V.~F. Mukhanov, {The Global Structure of the
  Inflationary Universe}, Int. J. Mod. Phys. A2 (1987) 561--591.
\newblock \href {https://doi.org/10.1142/S0217751X87000211}
  {\path{doi:10.1142/S0217751X87000211}}.

\bibitem{Guth:2000ka}
A.~H. Guth, {Inflation and eternal inflation}, Phys. Rept. 333 (2000) 555--574.
\newblock \href {http://arxiv.org/abs/astro-ph/0002156}
  {\path{arXiv:astro-ph/0002156}}, \href
  {https://doi.org/10.1016/S0370-1573(00)00037-5}
  {\path{doi:10.1016/S0370-1573(00)00037-5}}.

\bibitem{Guth:2007ng}
A.~H. Guth, {Eternal inflation and its implications}, J. Phys. A40 (2007)
  6811--6826.
\newblock \href {http://arxiv.org/abs/hep-th/0702178}
  {\path{arXiv:hep-th/0702178}}, \href
  {https://doi.org/10.1088/1751-8113/40/25/S25}
  {\path{doi:10.1088/1751-8113/40/25/S25}}.

\bibitem{Linde:2015edk}
A.~Linde, {A brief history of the multiverse}, Rept. Prog. Phys. 80~(2) (2017)
  022001.
\newblock \href {http://arxiv.org/abs/1512.01203} {\path{arXiv:1512.01203}},
  \href {https://doi.org/10.1088/1361-6633/aa50e4}
  {\path{doi:10.1088/1361-6633/aa50e4}}.

\bibitem{Barenboim:2016mmw}
G.~Barenboim, W.-I. Park, W.~H. Kinney, {Eternal Hilltop Inflation}, JCAP
  1605~(05) (2016) 030.
\newblock \href {http://arxiv.org/abs/1601.08140} {\path{arXiv:1601.08140}},
  \href {https://doi.org/10.1088/1475-7516/2016/05/030}
  {\path{doi:10.1088/1475-7516/2016/05/030}}.

\bibitem{Assadullahi:2016gkk}
H.~Assadullahi, H.~Firouzjahi, M.~Noorbala, V.~Vennin, D.~Wands, {Multiple
  Fields in Stochastic Inflation}, JCAP 1606~(06) (2016) 043.
\newblock \href {http://arxiv.org/abs/1604.04502} {\path{arXiv:1604.04502}},
  \href {https://doi.org/10.1088/1475-7516/2016/06/043}
  {\path{doi:10.1088/1475-7516/2016/06/043}}.

\bibitem{Starobinsky:1986fx}
A.~A. Starobinsky, {STOCHASTIC DE SITTER (INFLATIONARY) STAGE IN THE EARLY
  UNIVERSE}, Lect. Notes Phys. 246 (1986) 107--126.
\newblock \href {https://doi.org/10.1007/3-540-16452-9_6}
  {\path{doi:10.1007/3-540-16452-9_6}}.

\bibitem{Starobinsky:1994bd}
A.~A. Starobinsky, J.~Yokoyama, {Equilibrium state of a selfinteracting scalar
  field in the De Sitter background}, Phys. Rev. D50 (1994) 6357--6368.
\newblock \href {http://arxiv.org/abs/astro-ph/9407016}
  {\path{arXiv:astro-ph/9407016}}, \href
  {https://doi.org/10.1103/PhysRevD.50.6357}
  {\path{doi:10.1103/PhysRevD.50.6357}}.

\bibitem{Vennin:2015hra}
V.~Vennin, A.~A. Starobinsky, {Correlation Functions in Stochastic Inflation},
  Eur. Phys. J. C75 (2015) 413.
\newblock \href {http://arxiv.org/abs/1506.04732} {\path{arXiv:1506.04732}},
  \href {https://doi.org/10.1140/epjc/s10052-015-3643-y}
  {\path{doi:10.1140/epjc/s10052-015-3643-y}}.

\bibitem{Pattison:2017mbe}
C.~Pattison, V.~Vennin, H.~Assadullahi, D.~Wands, {Quantum diffusion during
  inflation and primordial black holes}, JCAP 1710~(10) (2017) 046.
\newblock \href {http://arxiv.org/abs/1707.00537} {\path{arXiv:1707.00537}},
  \href {https://doi.org/10.1088/1475-7516/2017/10/046}
  {\path{doi:10.1088/1475-7516/2017/10/046}}.

\bibitem{Noorbala:2018zlv}
M.~Noorbala, V.~Vennin, H.~Assadullahi, H.~Firouzjahi, D.~Wands, {Tunneling in
  Stochastic Inflation}[JCAP1809,no.09,032(2018)] (2018).
\newblock \href {http://arxiv.org/abs/1806.09634} {\path{arXiv:1806.09634}},
  \href {https://doi.org/10.1088/1475-7516/2018/09/032}
  {\path{doi:10.1088/1475-7516/2018/09/032}}.

\bibitem{Hawking:1981fz}
S.~W. Hawking, I.~G. Moss, {Supercooled Phase Transitions in the Very Early
  Universe}, Phys. Lett. 110B (1982) 35--38, [Adv. Ser. Astrophys.
  Cosmol.3,154(1987)].
\newblock \href {https://doi.org/10.1016/0370-2693(82)90946-7}
  {\path{doi:10.1016/0370-2693(82)90946-7}}.

\bibitem{Aghanim:2018eyx}
N.~Aghanim, et~al., {Planck 2018 results. VI. Cosmological parameters} (2018).
\newblock \href {http://arxiv.org/abs/1807.06209} {\path{arXiv:1807.06209}}.

\bibitem{Takahashi:2013cxa}
F.~Takahashi, {New inflation in supergravity after Planck and LHC}, Phys. Lett.
  B727 (2013) 21--26.
\newblock \href {http://arxiv.org/abs/1308.4212} {\path{arXiv:1308.4212}},
  \href {https://doi.org/10.1016/j.physletb.2013.10.026}
  {\path{doi:10.1016/j.physletb.2013.10.026}}.

\bibitem{Nakayama:2012dw}
K.~Nakayama, F.~Takahashi, {PeV-scale Supersymmetry from New Inflation}, JCAP
  1205 (2012) 035.
\newblock \href {http://arxiv.org/abs/1203.0323} {\path{arXiv:1203.0323}},
  \href {https://doi.org/10.1088/1475-7516/2012/05/035}
  {\path{doi:10.1088/1475-7516/2012/05/035}}.

\bibitem{Bunch:1978yq}
T.~S. Bunch, P.~C.~W. Davies, {Quantum Field Theory in de Sitter Space:
  Renormalization by Point Splitting}, Proc. Roy. Soc. Lond. A360 (1978)
  117--134.
\newblock \href {https://doi.org/10.1098/rspa.1978.0060}
  {\path{doi:10.1098/rspa.1978.0060}}.

\bibitem{Czerny:2014wza}
M.~Czerny, F.~Takahashi, {Multi-Natural Inflation}, Phys. Lett. B733 (2014)
  241--246.
\newblock \href {http://arxiv.org/abs/1401.5212} {\path{arXiv:1401.5212}},
  \href {https://doi.org/10.1016/j.physletb.2014.04.039}
  {\path{doi:10.1016/j.physletb.2014.04.039}}.

\bibitem{Czerny:2014xja}
M.~Czerny, T.~Higaki, F.~Takahashi, {Multi-Natural Inflation in Supergravity},
  JHEP 05 (2014) 144.
\newblock \href {http://arxiv.org/abs/1403.0410} {\path{arXiv:1403.0410}},
  \href {https://doi.org/10.1007/JHEP05(2014)144}
  {\path{doi:10.1007/JHEP05(2014)144}}.

\bibitem{Flauger:2009ab}
R.~Flauger, L.~McAllister, E.~Pajer, A.~Westphal, G.~Xu, {Oscillations in the
  CMB from Axion Monodromy Inflation}, JCAP 1006 (2010) 009.
\newblock \href {http://arxiv.org/abs/0907.2916} {\path{arXiv:0907.2916}},
  \href {https://doi.org/10.1088/1475-7516/2010/06/009}
  {\path{doi:10.1088/1475-7516/2010/06/009}}.

\bibitem{Kobayashi:2010pz}
T.~Kobayashi, F.~Takahashi, {Running Spectral Index from Inflation with
  Modulations}, JCAP 1101 (2011) 026.
\newblock \href {http://arxiv.org/abs/1011.3988} {\path{arXiv:1011.3988}},
  \href {https://doi.org/10.1088/1475-7516/2011/01/026}
  {\path{doi:10.1088/1475-7516/2011/01/026}}.

\bibitem{Takahashi:2013tj}
F.~Takahashi, {The Spectral Index and its Running in Axionic Curvaton}, JCAP
  1306 (2013) 013.
\newblock \href {http://arxiv.org/abs/1301.2834} {\path{arXiv:1301.2834}},
  \href {https://doi.org/10.1088/1475-7516/2013/06/013}
  {\path{doi:10.1088/1475-7516/2013/06/013}}.

\bibitem{Linde:1981mu}
A.~D. Linde, {A New Inflationary Universe Scenario: A Possible Solution of the
  Horizon, Flatness, Homogeneity, Isotropy and Primordial Monopole Problems},
  Phys. Lett. 108B (1982) 389--393, [Adv. Ser. Astrophys. Cosmol.3,149(1987)].
\newblock \href {https://doi.org/10.1016/0370-2693(82)91219-9}
  {\path{doi:10.1016/0370-2693(82)91219-9}}.

\bibitem{Albrecht:1982wi}
A.~Albrecht, P.~J. Steinhardt, {Cosmology for Grand Unified Theories with
  Radiatively Induced Symmetry Breaking}, Phys. Rev. Lett. 48 (1982)
  1220--1223, [Adv. Ser. Astrophys. Cosmol.3,158(1987)].
\newblock \href {https://doi.org/10.1103/PhysRevLett.48.1220}
  {\path{doi:10.1103/PhysRevLett.48.1220}}.

\bibitem{Guth:1980zm}
A.~H. Guth, {The Inflationary Universe: A Possible Solution to the Horizon and
  Flatness Problems}, Phys. Rev. D23 (1981) 347--356, [Adv. Ser. Astrophys.
  Cosmol.3,139(1987)].
\newblock \href {https://doi.org/10.1103/PhysRevD.23.347}
  {\path{doi:10.1103/PhysRevD.23.347}}.

\bibitem{Starobinsky:1980te}
A.~A. Starobinsky, {A New Type of Isotropic Cosmological Models Without
  Singularity}, Phys. Lett. B91 (1980) 99--102, [,771(1980)].
\newblock \href {https://doi.org/10.1016/0370-2693(80)90670-X}
  {\path{doi:10.1016/0370-2693(80)90670-X}}.

\bibitem{Sato:1980yn}
K.~Sato, {First Order Phase Transition of a Vacuum and Expansion of the
  Universe}, Mon. Not. Roy. Astron. Soc. 195 (1981) 467--479.

\end{thebibliography}
\end{document}